\documentclass[12pt,a4paper,twocolumn]{narms2}

\usepackage{graphicx}
\usepackage{subfigure} 
\usepackage{psfig}
\usepackage{timesmt}   
\usepackage{chikako}   

\makeatletter
\renewcommand{\section}{\@startsection%
{section}%
{1}%
{0mm}%
{- \baselineskip}%
{0.15\baselineskip}%
{\normalfont\normalsize}}%

\renewcommand{\subsection}{\@startsection
{subsection}%
{2}%
{0mm}%
{-\baselineskip}%
{0.15\baselineskip}%
{\normalfont\normalsize}}%
\makeatother

\begin{document}

\title{Dynamic and instability of submarine avalanches}
\author{\large{F. Malloggi \& J. Lanuza \& B.Andreotti \& E. Cl\'{e}ment.}\\{\em Physique et M\'{e}canique des Milieux H\'{e}t\'{e}rog\'{e}nes, Ecole Sup\'{e}rieure de Physique et de Chimie Industrielles}\\ {\em 10 rue Vauquelin, 75231 Paris Cedex 05, France.}
}

\date{}

\abstract{We perform a laboratory-scale experiment of submarine avalanches on a rough inclined plane. A sediment layer is prepared and thereafter tilted up to an angle lower than the spontaneous avalanche angle. The sediment is scrapped until an avalanche is triggered. Based on the stability diagram of the sediment layer, we investigate different structures for the avalanche front dynamics. First we see a straight front descending the slope, and then a transverse instability occurs. Eventually, a fingering instability shows up similar to rivulets appearing for a viscous fluid flowing down an incline. The mechanisms leading to this new instability and the wavelength selection are discussed.}


\maketitle
\frenchspacing   


\section{INTRODUCTION}

Avalanches are common situations implying dense sediment transport falling down a slope. It is well known that it is a phenomenon that may have practical dramatic implications and for this reason, is the object of intense studies ranging from laboratory scale experiments up to extensive field measurements. 
Here, we consider a laboratory situation modeling underwater avalanches triggered on a substrate that can be eroded. In this set up, the initial height of the sediment deposit as well as the plane slope are the two control parameters which allow an exploration of a large range of stable and metastable equilibrium conditions. A key issue is to understand the dynamics of matter waves falling down a slope(an avalanche) and to which extend such a wave is susceptible to amplify or die. Such a simple experiment study is also likely to provide strong tests that would help the development of reliable erosion/deposition theoretical models. 
Previously extensible studies on dry granular materials falling down a rough plane, have strongly contributed to a better understanding of the rheology of dense granular assemblies \shortcite{Pouliquen99}\shortcite{MiDi}. For erosive substrates, recent contributions have shown the existence of localized erosion waves in the form of triangular shape avalanches \shortcite{DD99}. Nevertheless, at the fundamental level, a complete understanding of all the physics involved in such a complex erosion/deposition process is still a challenging issue. Furthermore, the correspondence between dry grains avalanches and the underwater situation is largely unexplored.
The paper is organized as follows. Sec.2 de-scribes the experimental set-up. Sec.3 the different experimental results and Sec.4 gives a summary of the obtained results and bring some perspectives.

\section{EXPERIMENTAL SET-UP}
\subsection{The sediment layer}
We use alumina oxide powders of size $d=$ 15, 20, 30 and 40 $\mu m$. For such grain sizes it is usually difficult to obtain a non cohesive material. Under water, the aluminum oxide undergoes a dissolution releasing $OH^{-}$ ions. The pH value increases until the isoelectric point ($pH\approx9.3$). To avoid such a situation, we have shown that it is necessary to maintain pH values close to 4 by addition of hydrochloric acid \shortcite{Desset00} \shortcite{Daerr03}.

\subsection{The avalanche plane}
The grains are deposited on a $15 cm x 15 cm$ plexiglass substrate fixed at the bottom of a container that can be tilted at an angle value  ranging from 0 up top. The substrate was previously abraded in order to get a uniform roughness of about 1 m. At the beginning of the experiment, it is set to a horizontal position and a fixed mass of powder is poured and suspended by vigorous stirring. Then, a uniform sediment of height forms within 5 to 30 min depending on d.
\subsection{Triggering avalanches}

To initiate the avalanche fronts, we designed a "bulldozer" technique (inset of figure \ref{fig:stabdiag}) i.e. a vertical plate scraping the sediment at constant velocity $V_{p}$ chosen at a value about one half of the typical avalanching velocity $V_{a}$. In water $V_{a}$ is of the order of the Stokes velocity $V_{s}$:

$V_{s}=\frac{\Delta\rho}{\rho_{w}}\frac{g d^{2}}{\nu}$, where $\frac{\Delta\rho}{\rho_{w}}=3$
\\
is the density contrast between the grain and the water, $\nu$is the water kinematic viscosity and $g$ the gravity acceleration.

\subsection{Measurements}
Avalanche fronts are extracted using a CCD camera, coupled to a data acquisition card and computer performing a subsequent image analysis. A laser slicing technique allows the measurement of the front vertical shape. The lateral deviation of the laser sheet is proportional to the height $h$ of the front avalanche with the relation:
\\
$\Delta=\frac{h}{tan\alpha}$ where $\alpha$ is the incidence angle of the laser sheet.
\section{EXPERIMENTAL RESULTS}
\subsection{Stability diagram}

\begin{figure}[hh]
\begin{center}
\includegraphics[scale=0.2]{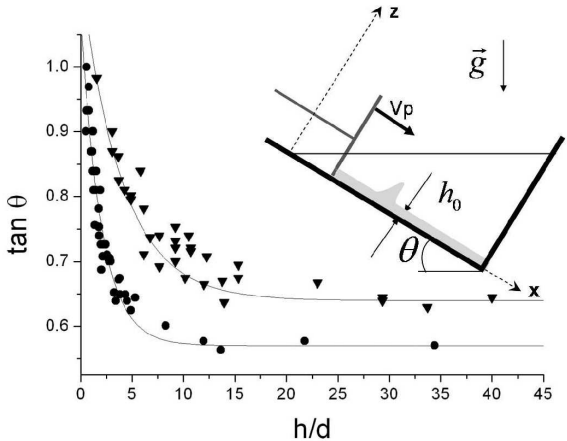}
\caption{\small  Stability diagram. Circles correspond to the thickness of the sediment left after an avalanche for a given angle, $\theta_{stop}(h)$ curve ($d$=30 and 40 $\mu m$). Triangles show the maximum stable height of sediment for a fixed angle, $\theta_{start}(h)$curve.}
\label{fig:stabdiag}
\end{center}
\end{figure}

For dry granular deposits, the stability of a static uniform layer $h$ lying on an inclined plate $\theta$ can be simply apprehended by a diagram with two branches $\theta_{start}(h)$ and $\theta_{stop}(h)$. For slopes above $\theta_{start}(h)$ the whole layer of height $h$ is unstable. The thickness $hstop(\theta)$ of the sediment left after an avalanche for a given angle $\theta$ yields the curve $\theta_{stop}(h)$ \shortcite{Pouliquen99}. Between the angles $\theta_{start}(h)$ and $\theta_{stop}(h)$,  we have a metastable situation where a finite perturbations of the substrate may propagate and amplify yielding an avalanching process \shortcite{DD99}. For depths typically larger than 10 grain sizes, the $\theta_{start}(h)$ curve reaches an asymptotic angle limit $\theta_{a}$, the avalanche angle of the granular pile and the $\theta_{stop}(h)$ curve reaches the asymptotic value $\theta_{r}$ the repose angle.  Actually, for our underwater substrate we find a similar stability diagram. Figure \ref{fig:stabdiag} shows this diagram for several sizes of aluminum oxide powder.

\begin{figure}
\begin{center}
\includegraphics[scale=0.3]{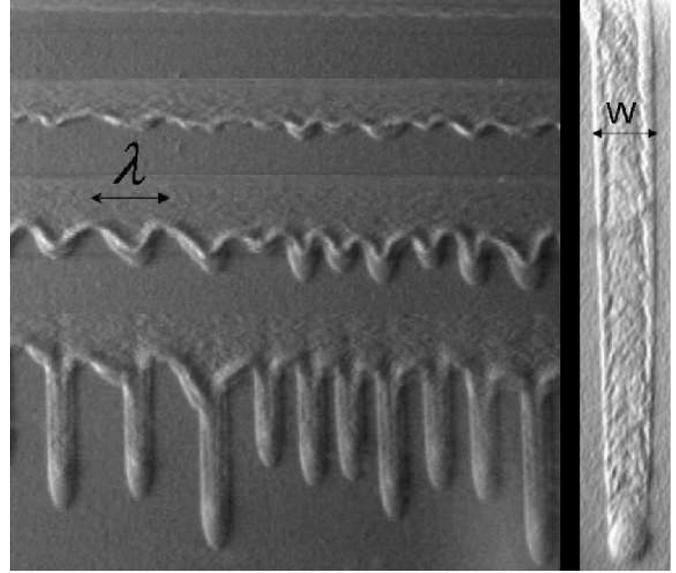}
\caption{\small  a)Typical evolution of the front during an avalanche. b)Zoom on a finger.}
\label{fig:evolfront}
\end{center}
\end{figure}

\subsection{Instability}
\begin{figure}[hh]
\begin{center}
\includegraphics[scale=0.4]{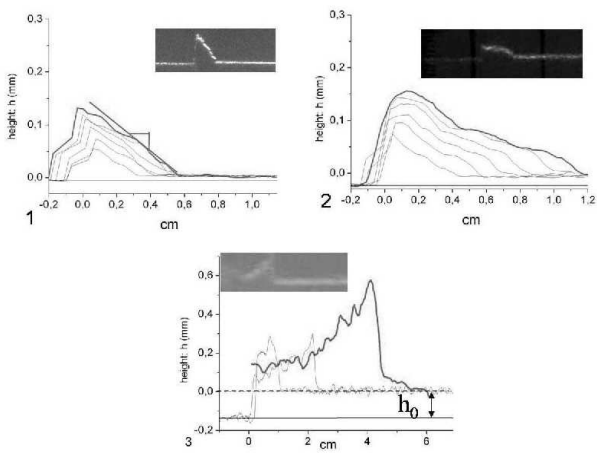}

\caption{\small  Temporal evolution of the front avalanche shape visualized with a laser slice. 1. Beginning of the experiment, we scrap matter and the front keeps a triangle shape. 2. Then the sediment spreads, with a formation of a tong of material. 3. Finally an avalanche is triggered with this characteristic vertical profile and propagates downhill (note here that the height scale is at least twice larger than previous scales).}
\label{fig:shape}
\end{center}
\end{figure}

Now we describe the results of avalanche propagation experiments. For small angles, $\theta<\theta_{r}$ the sediment is stable as an avalanche front cannot propagate autonomously. For intermediate angles $\theta_{r}<\theta<\theta_{a}$, we could produce transversally stable avalanche fronts. In the case where the initial sediment height $h=h_{stop}(\theta )$, we observed neutral waves that could propagate autonomously through the whole set-up. An extensive study of such fronts will be presented elsewhere. Here we rather concentrate on the description of a transverse instability observed for avalanche fronts produced in the situation where $\theta>\theta_{a}$.
On figure \ref{fig:evolfront} we show the typical evolution of such a front exhibiting two typical features: first, a transverse instability and then, a fingering instability. Figure \ref{fig:shape} shows the detailed evolution of the front produced by the bulldozer: starting with a triangular shape, then a tongue spreading on the substrate and finally a shark teeth shape. At the rear of the avalanche, a height $h_{stop}(\theta )$ is found as expected.

\subsection{Coarsening process}

\begin{figure}[hh]
\begin{center}
\includegraphics[scale=0.6]{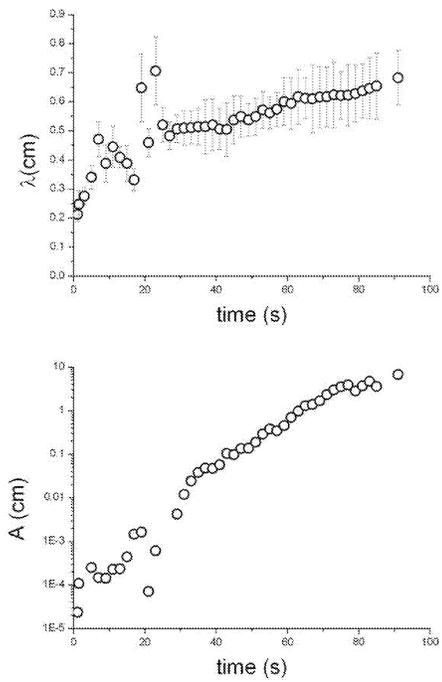}
\caption{\small  Temporal evolution of $\lambda$ in a lin-lin plot and the amplitude $A$ in a log-lin plot.}
\label{fig:coars}
\end{center}
\end{figure}

Now we present systematic measurements of the front shape that allows determining the transverse unstable wavelength as a function of time. By image processing, the shape of the avalanche front $x(y,t)$ is extracted and we compute the front line auto correlation function:
\\
$c(y_{0},t)=<x(y+y_{0},t)x(y,t)>-<x(y,t)>^{2}$
\par
The first maximum of this function, $\lambda (t)$, with a correlation value $C(y0,t) = A^{2}(t)$, represents the dominant transverse modulation length. The growth of $\lambda (t)$ and $A(t)$,  is characteristic of a coarsening process driving the front dynamic evolution: initially small undulations give rise to larger structures. This process is interrupted by the onset of the fingering process. On figure \ref{fig:coars}, we display the wavelength evolution:$\lambda (t)$ and the corresponding amplitude increase $A(t)$.
\par
\begin{figure}[hh]
\begin{center}
\includegraphics[scale=0.3]{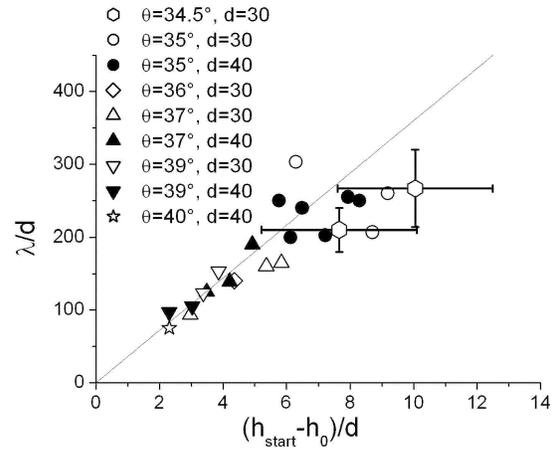}
\caption{\small  Maximal wavelength $\lambda_{max}$ as a function of
$(h_{start}(\theta ))-h_{0}/d$. The straight line is $y= 36 x$.
}
\label{fig:lambda}
\end{center}
\end{figure}

 
We observe a factor about 3 between the first wavelength detected and maximum value  $\lambda_{max}$. This maximal wavelength  $\lambda_{max}/d$ is plotted on figure \ref{fig:lambda}, as a function of $(h_{start}(\theta ))-h_{0}/d$. Interestingly, for all wavelength measured, we observe a data collapse with:
\\
$\lambda_{max}\approx A(h_{start}(\theta)-h_{0})/d$ where$A\approx36$

\par
A noticeable scattering between the data points still remains, especially of the larger wave lengths.  But one should notice first, that the error bars are large since measurements of $h_{start}$ are very noisy.  Second, for all data, the avalanche geometry may vary (loosing or gaining matter) since the substrate height is not exclusively at a value $h=h_{stop}$. Such a situation where the avalanche shape changes slowly could influence somehow the instability mechanism.

\subsection{Fingers}
The presence of a subsequent fingering instability is quite fascinating feature of this avalanching process. It bears many similarities with the fingering patterns observed for thin viscous layers flowing on a plate \shortcite{Huppert82} but in our case, no interplay between viscosity and surface tension can be claimed as a selection mechanism. Here, the fingering front stems from the onset of localized propagating wave solutions in the erosion/deposition process.
\begin{figure}[hh]
\begin{center}
\includegraphics[scale=0.7]{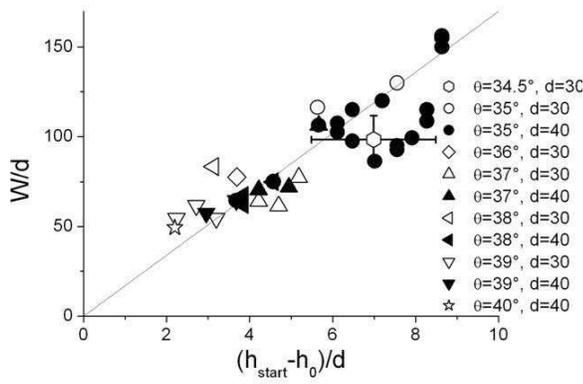}
\caption{\small  . Relative fingers width $W/d$ as a function of 
$h_{start}(\theta)-h_{0})/d$. The straight line is $y= 17x$.}
\label{fig:W}
\end{center}
\end{figure}
\par
The finger width $W(\theta,h_{0})$ is also found to be sensitive to the slope angle and, as for the destabilization wavelength $\lambda_{max}$, we propose a rescaling: $W\approx A(h_{start}(\theta)-h_{0})/d$ with $B\approx17$ (see fig\ref{fig:W})

\par
A striking feature of the fingers hence obtained is the presence of levees on the side and of a drop like head at the front (see fig 2), which is reminiscent of many natural patterns obtained in debris or mud flows. Nevertheless, the formation mechanism here is different from the segregation induced fingering observed by Pouliquen et al. \shortcite{Pouliquen99b}.

\section{CONCLUSION AND PERSPECTIVE }

In summary, we have reported here an experimental work of submarine avalanches. The stability diagram of the sediment layer is similar to what is observed for a dry granular deposit. We evidence a regime where the evolution of the avalanche front shows a transverse instability characterized by a coarsening process and followed by a fingering instability. We systematically extract the wavelength and show that it is quite sensitive to the slope angle. We tentatively propose a rescaling relation, for the wavelength and the width of the observed fingers, as a function of the relative height between the sediment thickness and the maximum height that can be sustained for a given slope. The fingers morphology (levees, drops) is also an important observation of the experiment.
\par
In water, as we operate, the pore pressure varia-tions and water back flows could play a crucial role in the avalanche dynamics. Nevertheless preliminary results obtained on dry avalanches, seem to bear a similar phenomenology which indicates that the pre-sent instability is more likely to be related to fundamental aspects of erosion/deposition dynamics.
\par
We thank A Daerr, S. Douady and O. Pouliquen for many useful discussions and comments.

\newpage

\bibliographystyle{chikako}      
\bibliography{maldai2} 

\end{document}